\documentclass{vgtc}                          % final (conference style)
\let\ifpdf\relax
\usepackage{mathptmx}
\usepackage{graphicx}
\usepackage{times}
\usepackage{lipsum}
\usepackage{subfigure}
\usepackage{balance}  % to better equalize the last page
\usepackage{graphics} % for EPS, load graphicx instead
\usepackage{times}    % comment if you want LaTeX's default font
\usepackage{url}      % llt: nicely formatted URLs
\usepackage{placeins}
\usepackage{xspace}
\usepackage{paralist}
\usepackage{color} 
\usepackage{algorithm}
\usepackage[noend]{algpseudocode}

\newcommand{\review}[1]{#1}

\def\includeFigures{}

\onlineid{167}

\vgtccategory{Research}

\vgtcinsertpkg

\title{Adaptive User Perspective Rendering for Handheld Augmented Reality}

\author{
    Peter Mohr\textsuperscript{1}\thanks{e-mail: mohr@icg.tugraz.at}\\ %
\and Markus Tatzgern\textsuperscript{2}\thanks{e-mail: markus.tatzgern@fh-salzburg.ac.at}\\ %
\and Jens Grubert\textsuperscript{3}\thanks{e-mail: jg@jensgrubert.de}\\ %
\and Dieter Schmalstieg\textsuperscript{1}\thanks{e-mail: schmalstieg@icg.tugraz.at}\\ %
\and Denis Kalkofen\textsuperscript{1}\thanks{e-mail: kalkofen@icg.tugraz.at}\\ %
}
\affiliation{
\centering
\vspace{-0.2cm}
\scriptsize \textsuperscript{1}Graz University of Technology
\hspace{1cm}
\scriptsize \textsuperscript{2}Salzburg University of Applied Sciences
\hspace{1cm}
\scriptsize \textsuperscript{3}Coburg University
}

\teaser{
  \vspace{-0.2cm}
    \centering    
	\subfigure[Device Perspective Rendering]
	{\includegraphics[width=0.45\textwidth]{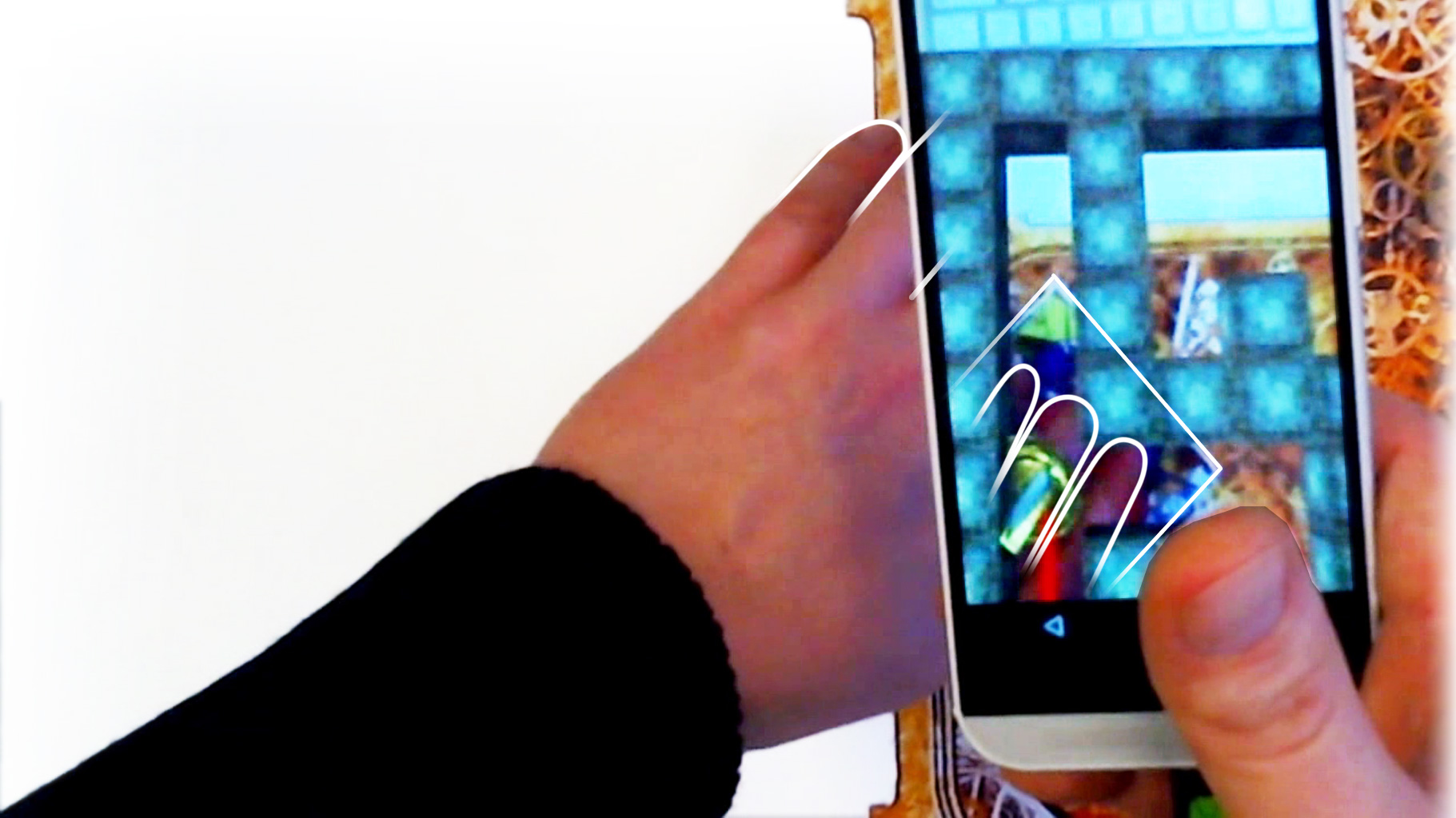}}
	\hfill 
	\subfigure[User Perspective Rendering]
	{\includegraphics[width=0.45\textwidth]{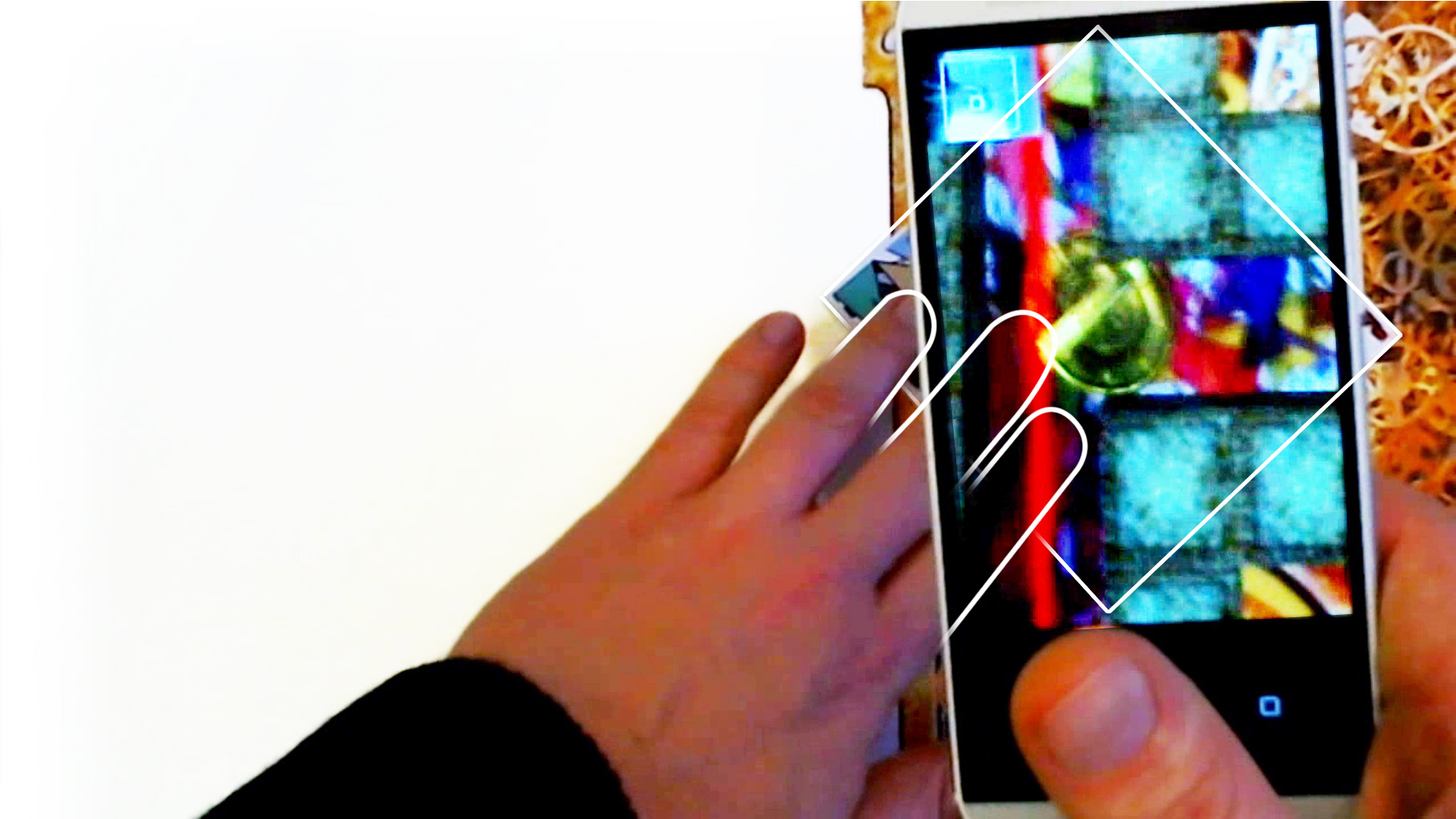}}
    \vspace{-0.3cm}
    \caption{Hand Interaction in Device and User Perspective Augmented Reality. (a) Device perspective rendering directly augments the video stream of the handheld device. Objects outside and inside the augmentations appear disconnected. Notice the hand inside the AR device. (b) User perspective rendering estimates the user's head pose in order to adapt the AR rendering as seen from the head position. Therefore, objects outside and inside the AR display visually connect. Notice the fingers visually connect to the hand of the user.}
    \label{fig:teaser}
    \vspace{-.5cm}
}

\abstract{
Handheld Augmented Reality commonly implements some variant of magic lens rendering, which turns only a fraction of the user’s real environment into AR while the rest of the environment remains unaffected. Since handheld AR devices are commonly equipped with video see-through capabilities, AR magic lens applications often suffer from spatial distortions, because the AR environment is presented from the perspective of the camera of the mobile device. Recent approaches counteract this distortion based on estimations of the user’s head position, rendering the scene from the user's perspective. To this end, approaches usually apply face-tracking algorithms on the front camera of the mobile device. However, this demands high computational resources and therefore commonly affects the performance of the application beyond the already high computational load of AR applications. In this paper, we present a method to reduce the computational demands for user perspective rendering by applying lightweight optical flow tracking and an estimation of the user’s motion before head tracking is started. We demonstrate the suitability of our approach for computationally limited mobile devices and we compare it to device perspective rendering, to head tracked user perspective rendering, as well as to fixed point of view user perspective rendering.
} % end of abstract

\CCScatlist{ 
  \CCScat{H.5.1}{Information Interfaces and Presentation}{Multimedia Information Systems - Artificial, augmented and virtual realities};
}


\begin{document}


%\firstsection{Introduction}

\maketitle
\newpage

\ifdefined\includeFigures
\begin{figure*}[t!]
\centering
	\subfigure[Device Perspective Rendering]
	{\includegraphics[width=0.33\textwidth]{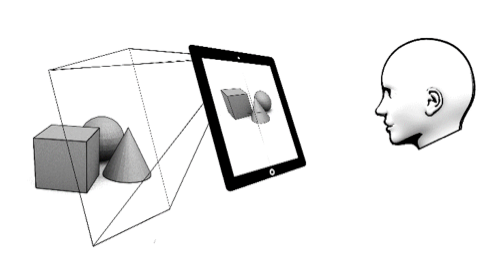}}	
	\subfigure[User Perspective Rendering]
	{\includegraphics[width=0.33\textwidth]{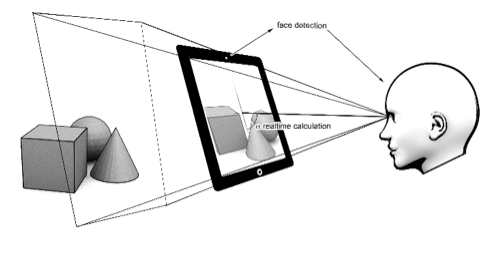}}		
	\subfigure[Approximated User Perspective Rendering]
	{\includegraphics[width=0.3\textwidth]{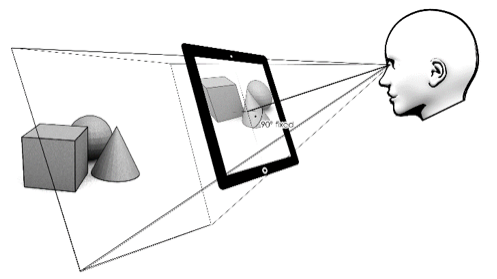}}
	\vspace{-0.25cm}
	\caption{Traditional approaches to magic lens rendering for handheld AR. 
	    (a) Device perspective rendering provides augmentations from the point of view of the camera. 
	    (b) User perspective rendering uses 3D head tracking to provide augmentation from the user's point of view. 
	 (c) Fixed point of view user perspective rendering does not require 3D head tracking. Instead it assumes a static spatial relationship between the user's head and the display surface.}
	 \vspace{-0.3cm}
\label{fig:rendermodes}
\end{figure*}
\fi

\section{Introduction} 
The computational capabilities of the current generation of smartphones and tablets enable handheld Augmented Reality (AR) applications for %many aspects of life
various application domains. For example, handheld AR has been successfully demonstrated to provide visual support in maintenance~\cite{Mohr:2015:RTD} and construction site monitoring~\cite{Zollmann:2012:IOD}, to interact in AR games ~\cite{Andrukaniec:2013:OAR} or tourist maps~\cite{grubert2015utility}, to annotate the real environment~\cite{Tatzgern:2014:HL}, and in many more situations~\cite{schmalstieg2015augmented}.

 %Therefore, most handheld AR applications implement some form of magic lens rendering~\cite{Bier:1993:TML}. Magic lenses turn only a fraction of the user’s real environment into AR, while the rest of the environment remains unaffected. This is usually implemented in a video see-through AR display, by merging computer graphics with the video data from the camera feed of the mobile device. 

%Since handheld AR is commonly implemented on mobile hardware, AR applications often have to deal with small-sized displays. %Therefore, most handheld AR applications implement some form of magic lens rendering~\cite{Bier:1993:TML}. Magic lenses turn only a fraction of the user’s real environment into AR, while the rest of the environment remains unaffected. This is usually implemented in a video see-through AR display, by merging computer graphics with the video data from the camera feed of the mobile device. 
Handheld AR typically employs smartphone- or tablet-sized screen formats. The commonly applied Magic Lens metaphor, turns only a fraction of the user’s real field-of-view into an augmented scene, while the rest of the environment remains unaffected. Furthermore, AR magic lens applications on video see-through displays often suffer from spatial distortions, because the AR environment is presented from the perspective of the camera of the device. The camera is usually located in a corner on the back of the device and captures the scene with a camera-dependent field of view. This is commonly defined as device-perspective rendering (DPR). The mismatch between the camera's and the user's point and field of view results in mismatching visualizations inside and outside of the magic lens (and is also called dual-view problem~\cite{vcopivc2014use}). The visualization inside the magic lens depends on the camera parameters while the user's perception outside is based on his or her natural vision (Figure\ref{fig:rendermodes}(a)). This misalignment is especially confusing when virtual content inside the magic lens has to be visually connected to real objects outside the magic lens, and during interactions with the AR environment through the magic lens, i.e., when the user sees its own hand inside and outside the lens. 

This is demonstrated in Figure~\ref{fig:teaser} (a), showing a user interacting with an AR game through the magic lens of the handheld device. In the game, the user has to rotate a small physical marker in order to align a virtual mirror with a laser beam to control reflections of the beam. The interaction requires grasping the physical marker and subsequently rotating it. While the rotations may be performed entirely inside the AR magic lens, manipulating the marker involves moving the hand from the outside into the view inside the magic lens. When rendering the AR scene from the camera’s point of view (Figure~\ref{fig:teaser}(a)) the visual mismatch between parts of the hand, which are outside, and other parts of the hand which are inside, make precisely grasping the marker difficult. User perspective rendering (UPR) has been proposed to overcome this problem~\cite{barivcevic2012hand}. It aligns the AR view inside the magic lens to the view of the user outside of the magic lens, to create the illusion of looking through a transparent glass frame (Figure\ref{fig:rendermodes}(b)). In fact, it has been shown that users, who were never exposed to handheld AR before, expect UPR as the default view ~\cite{vcopivc2014use}.
Current approaches targeting  typically implement UPR by computing the user’s head position in each frame before they align the AR view based respectively. Figure~\ref{fig:teaser}(b) shows the AR view with UPR. Notice that the fingers inside the magic lens visually connect to the hand outside. This potentially makes selection tasks easier~\cite{barivcevic2012hand}. 

Implementations of UPR have often been explored using head tracking systems in laboratory setups. They either include stationary external camera tracking~\cite{samini2014perspective} or additional hardware setups such as depth sensors~\cite{barivcevic2012hand}. On mobile devices, UPR has been implemented using 3D face tracking based on the video feed of front camera of modern mobile devices~\cite{grubert2014towards}. While this approach works in theory, implementations suffer from the computational demands of the additional head tracking. Based on our experience, this results in an overall low performance of the application, which can also be traced down to the devices' entering thermal throttling mode to prevent overheating. %As a result, mobile UPR applications are unusable. 
As a result, the usability of UPR applications can be substantially be reduced in real-world settings. A computationally less demanding alternative for mobile devices has been proposed by Pucihar et al.~\cite{vcopivc2013evaluating}. Instead of tracking the user’s head pose the authors manually measure the distance of the head to the device once at the beginning of the application, and they assume the user looking perpendicular through the center of device over the entirety of the application (Figure\ref{fig:rendermodes}(c)). This approach is called \textit{Fixed Point of View} user perspective rendering (FUPR). It avoids the computational effort required to continuously track the users head pose. 

However, FUPR fails to generate user perspective graphics for large interaction spaces. For example, Figure 3(a) shows a maintenance scenario, which requires to touch switches and buttons at the top and at the bottom of a large electric cabinet. In such a scenario, the spatial relationship between the user’s head position and the handheld device frequently changes, which in turn will eventually render FUPR ineffective. 

In this paper, we combine the resource-friendly approach of FUPR with continuously effective UPR. We achieve this by adding a lightweight Kanade-Lucas-Tomasi (KLT) tracker~\cite{Shi1994,Lucas:1981:IIR} to the head-tracking pipeline of traditional UPR systems. We use KLT-tracking to estimate head motion in image space, which we use to subsequently decide whether the parameters of FUPR need to be refreshed. Since simple thresholding of user motion will eventually introduce a certain amount of error, we present the idea of dual thresholding which incorporates temporal and spatial thresholding in order to derive a more precise 3D head pose when head motion stops. By automatically updating the parameters based on a 3D head tracker, we furthermore do not require any manual initialization. Since our approach incorporates less updates of the head pose, it is also more stable and more robust in environments where visual tracking is difficult. 

In the following we describe our system architecture and we report performance measurements as well as the results of a user experiment comparing DPR, UPR, FUPR to our approach of adaptive user perspective rendering (AAUPR). \review{Notice, since the approach of Tomioka et al. \cite{tomioka2013approximated} is called \textit{Approximated User Perspective Rendering} (AUPR) we choose AAUPR for our technique.} %in order to avoid confusing acronyms.}

\ifdefined\includeFigures
\begin{figure*}[t!]
\centering
    \subfigure[]
	{\includegraphics[width=0.26\textwidth]{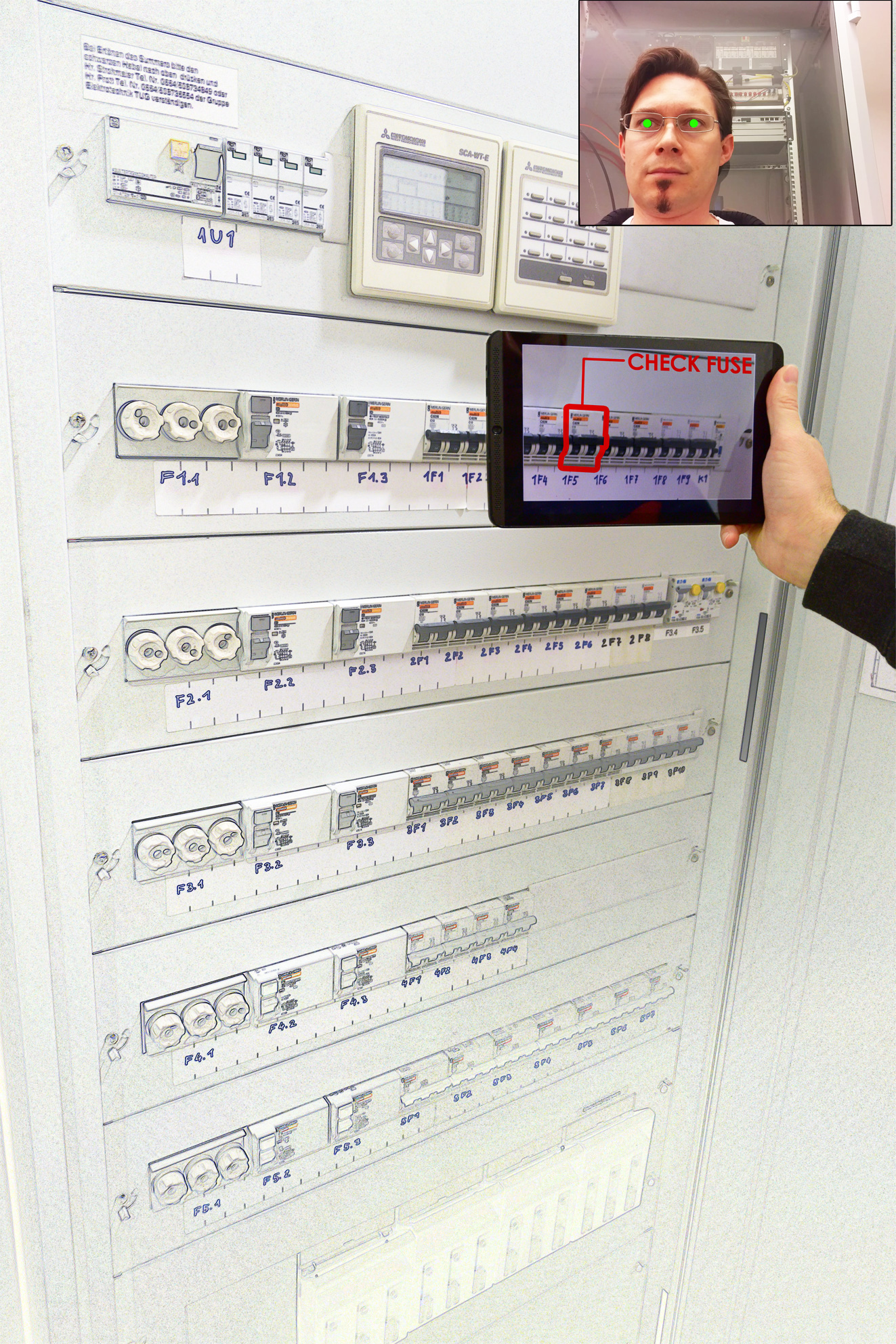}}	
	\subfigure[]
	{\includegraphics[width=0.455\textwidth]{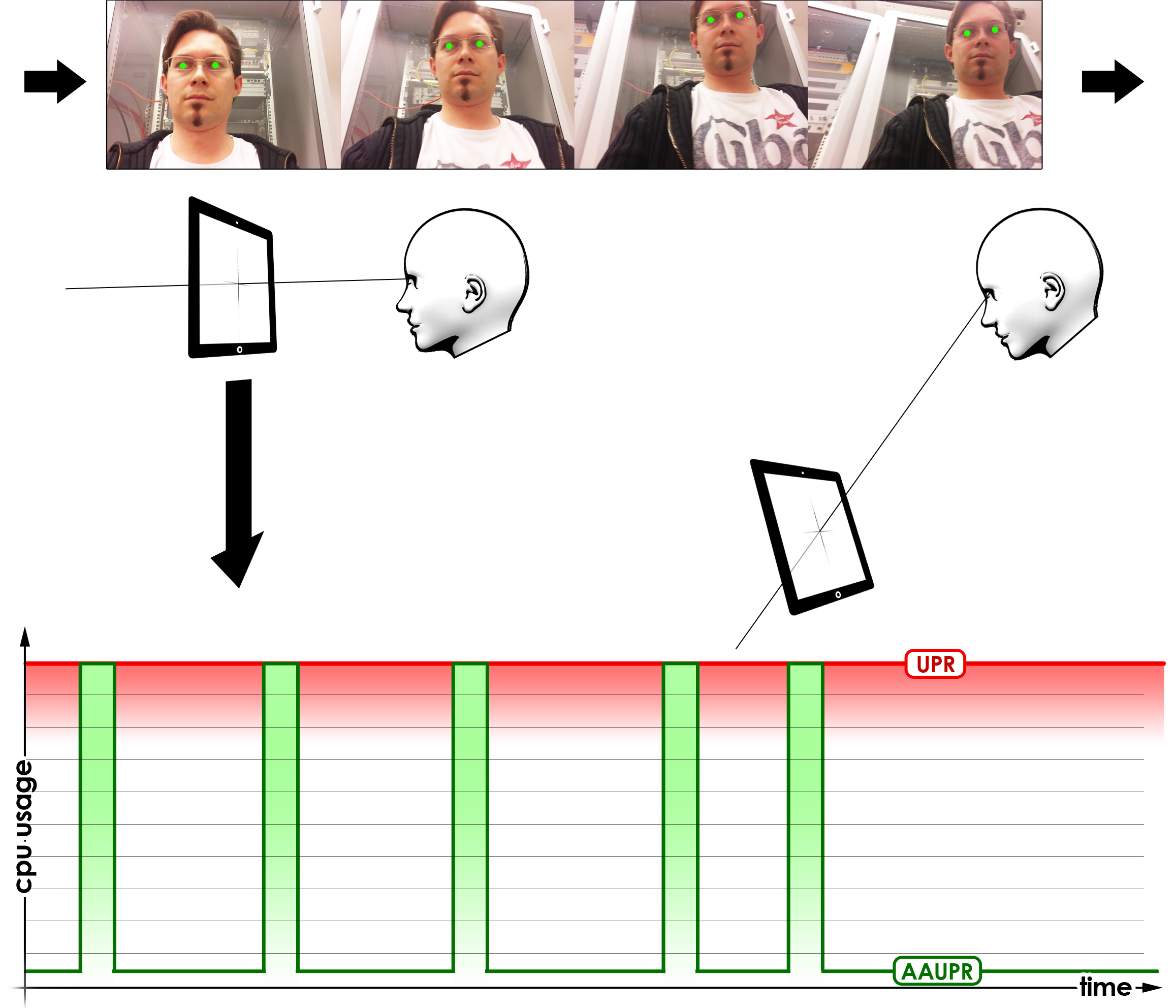}}		
	\subfigure[]
	{\includegraphics[width=0.26\textwidth]{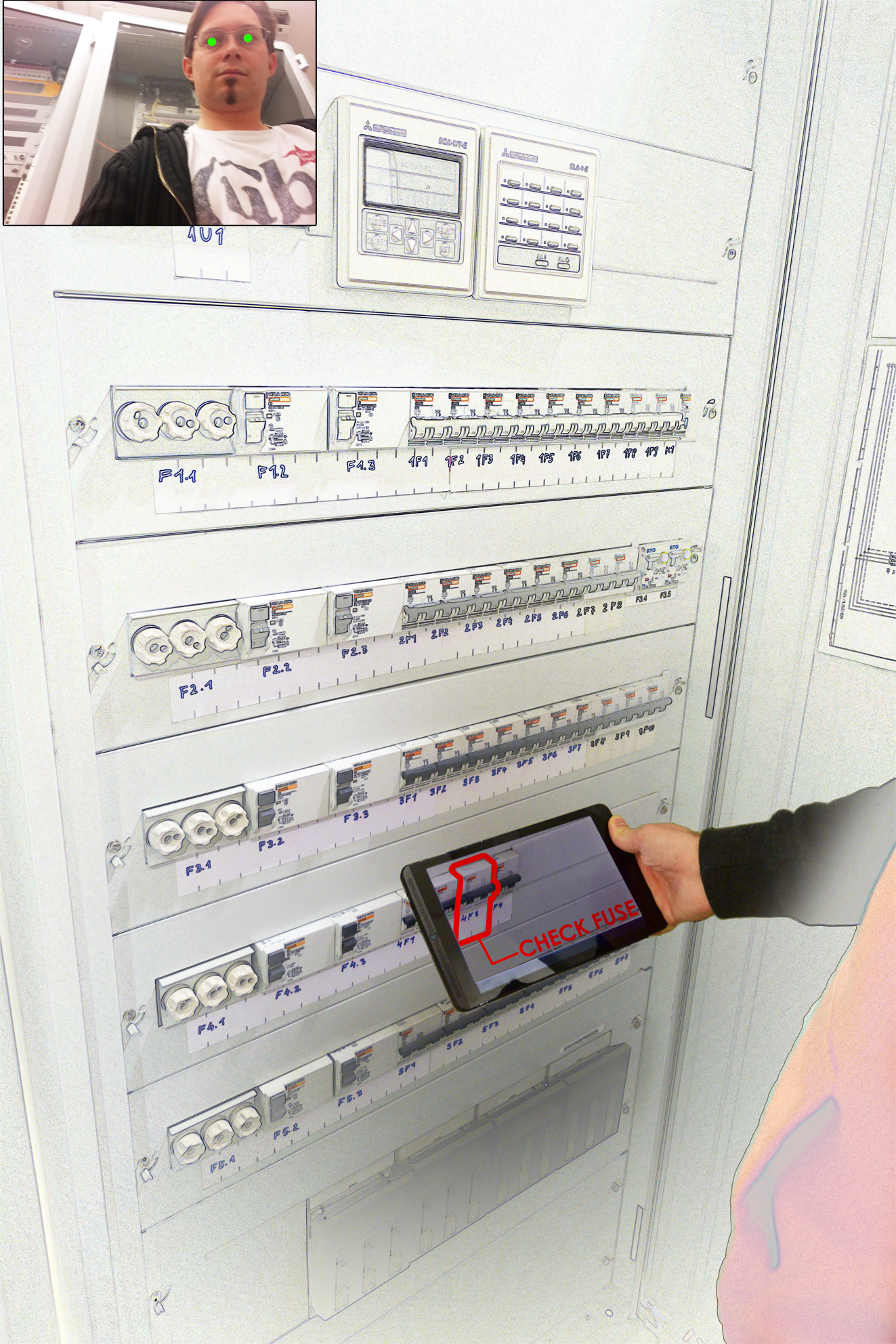}}
	\vspace{-0.3cm}
	\caption{System overview. In a typical usage scenario the user moves the handheld device from one position (a) to another (c) to view different AR instructions. The resulting transition of the user's head pose, in relation to the device, is illustrated in the upper row of (b) showing single images from the front camera of the mobile device. The diagram in (b) shows a symbolic graph of the CPU usage during our approach (AAUPR) compared to UPR. During user motion the head position is updated depending on the current threshold value. Once the user has moved to the desired point of view, the head pose is refined for this position (last peak in the graph).}
	%{\includegraphics[width=0.995\textwidth]{img/overview3.jpg}}	
	%\caption{System Overview. (a) (b)}
\label{fig:overview}
\vspace{-0.3cm}
\end{figure*}
\fi

%%%%%%%%%%%%%%%%%%%%%%%%
%%% RELATED WORK 
%%%%%%%%%%%%%%%%%%%%%%%%
\section{Related Work}
%UPR approaches have been primarily invesitgated vor video see-through handheld Augmented Reality systems such as smartphones or tablets.

Baricevic et al. evaluated the effects of display size on UPR and found that, using a simulation, a tablet-sized display allows for significantly faster performance of a selection task compared to a handheld display and that UPR outperformed DPR for a selection task\cite{barivcevic2012hand}. They also prototyped a UPR system with geometric reconstruction using a Kinect for reconstruction of the physical surrounding and a Wiimote for head tracking. In follow up works, they proposed to replace an active depth sensor by gradient domain image-based rendering method combined with semi-dense stereo matching \cite{barivcevic2014user, barivcevic2016user}. Other authors also employ depth-sensors for scene reconstruction in UPR \cite{Unuma2014SMA}.

Tomioka et al. and Hill et al. proposed an UPR implementation through transforming the back-facing camera image using homographies  \cite{tomioka2013approximated, hill2011virtual}. Samini et al. investigated UPR when using an external outside-in tracking system for spatial registration and proposed a geometric correction scheme for introduced registration errors \cite{samini2014perspective}.
 
Pucihar et al. investigated fixed point of view UPR (FUPR) versus DPR in a target acquisition task with and without scene continuity across device boundaries \cite{vcopivc2013evaluating}. Specifically, they assume that the user's face is in a fixed and predetermined position while interacting with the system They found that most users who never experienced handheld AR before actually expected UPR as the default mode of presentation. The study also indicated that UPR outperformed DPR in terms of accuracy, task completion time, subjective workload and preference. They later extended their investigations to specifically study the use of surrounding visual context in a map navigation task \cite{vcopivc2014use} and to sketching applications \cite{pucihar2015dual}. Pucihar et al. also proposed a specific variation of UPR, called contact-view, which allows pseudo transparent rendering of documents when a smartphone lies directly on the document \cite{pucihar2014poster1, pucihar2014poster2}, achieving similar effects compared to proprietary solutions using transparent displays \cite{hincapie2014car, hincapie2014tpad}.
    
Grubert et al. employed UPR on mobile devices by combining head tracking using the built-in front-facing RGB camera for head tracking and natural feature-based tracking of the AR device using back-facing RGB camera~\cite{grubert2014towards}. Rececently, Samini et al. compared UPR and DPR for a find-and-select and a 3D object manipulation task \cite{Samini2016}. While they found DPR to outperform UPR in terms of task completion time for the find-and-select task, both approaches where on par for an object manipulation task, and UPR was preferred by users.

Our approach is specifically targeting mobile devices which offer limited computational resources. Therefore, we have positioned our system between fully dynamic, but resource intensive UPR approaches relying on constant face tracking and the fixed point of view UPR approach (FUPR) of Pucihar et al. which is only applicable in constrained application scenarios.% such as looking straight onto a plane parallel to the device.

%\todo{@all 1-2 more sentences how our work differs from RW}
\ifdefined\includeFigures
\begin{figure*}[t!]
\centering
	%%%% 
	%%\subfigure[Application]
	{\includegraphics[width=0.995\textwidth]{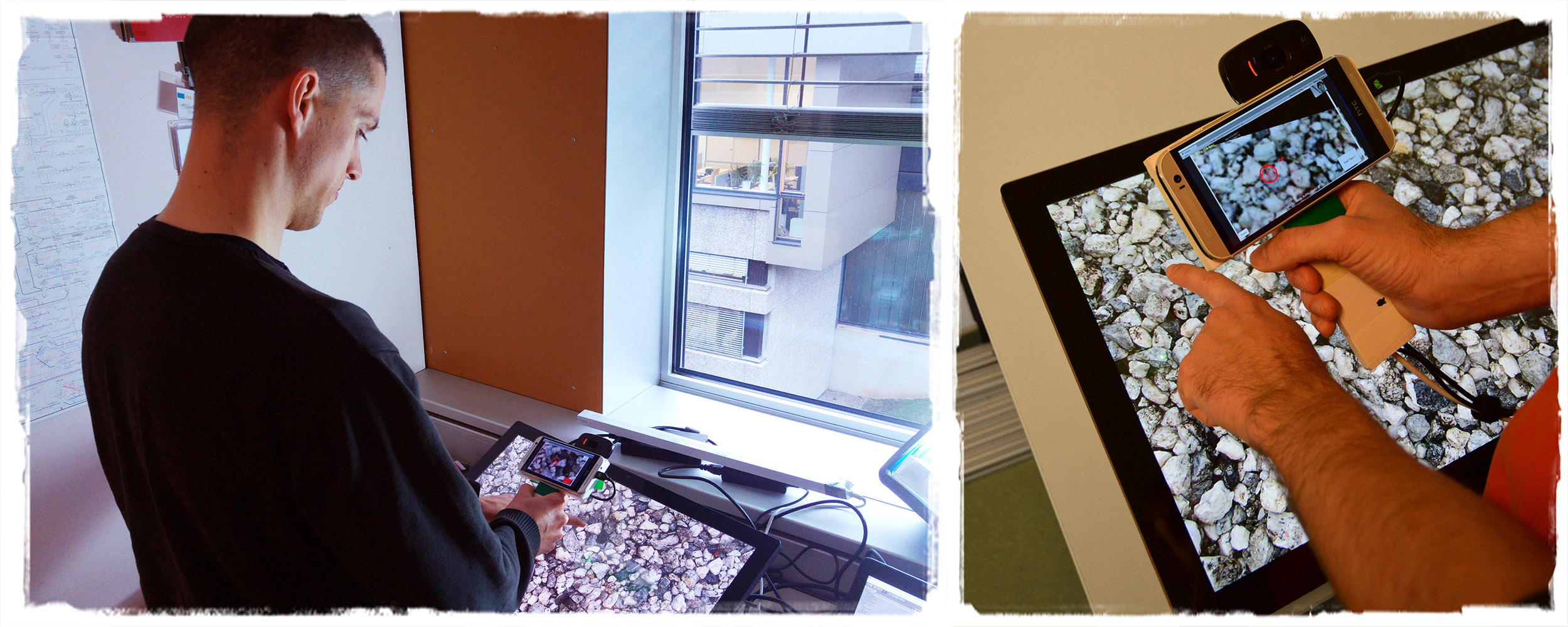}}
	\vspace{-0.3cm}
	\caption{User Experiment.}
	\vspace{-0.2cm}
\label{fig:experiment}
\end{figure*}
\fi

%%%%%%%%%%%%%%%%%%%%%%%%
%%% System
%%%%%%%%%%%%%%%%%%%%%%%%
%\newpage
%\vspace{-0.2cm}
\section{System}
%A number of research prototypes successfully demonstrate the implementation of user perspective rendering on high-end laboratory setups [refHoellerer]. In contrast, our system is designed to provide user perspective rendering specifically on computationally limited mobile devices. 
The main insight used for the design of our system is that updates on the user’s head pose are only necessary after a certain amount of motion and that small head motion can be ignored. Since FUPR assumes no head motion at all, it cannot support scenarios that require a large interaction space. Figure \ref{fig:overview} shows an example application where a maintenance worker is receiving visual instructions on a handheld AR device which include pressing buttons on a large electric cabinet. In such a scenario, the user has to correctly perceive instructions both, in the right top corner as well as in the corner on bottom left of the cabinet. Such large distances between points in space which need augmentations cannot be handled by FUPR systems because the calibration of the user’s spatial relationship to the display cannot hold.  

However, based on the user experiment provided by Pucihar et al.~\cite{vcopivc2013evaluating} we assume FUPR provides effective UPR for small to no changes of the user’s head position relative to the device. Therefore, we efficiently compute user perspective graphics by adding a low cost tracker to estimate user motion before we start expensive head tracking for motion larger than a certain threshold distance. %This approach can be seen as adaptive approximated user perspective rendering. 
In the following we explain the main components of our system: (1) user perspective rendering on mobile devices, (2) efficient motion estimation, and (3) dual thresholding.

%\noindent
\textbf{User Perspective Rendering on Mobile Devices.}
Traditional UPR requires estimating the user's 3D head position and the 6 degrees of freedom pose of the mobile device at every frame. Implementations on modern mobile smartphones derive the head pose using a 3D face tracker. The face tracker is usually applied to the video stream of the front camera while AR scene tracking is performed on the video stream of the back camera~\cite{grubert2014towards}. 

For device tracking we use natural feature tracking. In our current prototype we estimate the pose of the device using PTC's Vuforia SDK~\footnote{https://www.vuforia.com}. For 3D head tracking we use a combination of a 2D deformable FaceTracker~\cite{SaragihLC09} and a subsequent 3D model, similar to the approach of Grubert et al~\cite{grubert2014towards}.% which provides real time deformable face tracking and head pose estimation. %\todo{soll ich mehr schreiben?}

%\noindent
\textbf{Motion Estimation.}
We derive an estimate of the user's motion in order to handle updates of the 3D head tracker. While 3D head tracking is  expensive we aim at a low cost motion estimation. Therefore, we estimate motion in image space only. Our prototype uses KLT-tracking of few very distinctive features. We use the face tracker which was applied during the last 3D head pose estimation in order to find the points which describe the eyes of the user. Motion estimation is subsequently performed on these two points only. Whenever KLT-tracking fails we start full 3D head tracking to update UPR and to re-initialize the motion estimation.

\textbf{Dual Thresholding.}
After estimating the user's motion in image-space we apply simple thresholding to decide if an update of the user's 3D head pose is necessary. However, simple thresholding introduces an error relative to the size of the current threshold value. Furthermore, since we apply thresholding in image space the error increases for distant head poses. 

In order to reduce the error during interaction, we combine spatial with temporal thresholding. We assume that the spatial relationship between the AR display and the 3D head pose of the user mostly changes during scene exploration or while moving the display from one point of interaction to the next point. However, during interaction with the scene the 3D head pose usually stays rather steady. Therefore, we reduce the threshold over time and re-initialize it to its maximal value each time we estimate the 3D head pose using the head tracker. This approach allows us to provide regular updates of the 3D head pose during scene exploration, while also ensuring a precise 3D head pose during interaction (assuming stable head poses during interaction). \review{This approach is outlined in Algorithm \ref{alg::update} and illustrated in Figure~\ref{fig:overview}.}

\begin{algorithm}
%\begin{lstlisting}
\caption{Dual Thresholding}\label{alg::update}
\begin{algorithmic}[1]
\State $E \gets |PosEyeCalc - PosEyeFlow|$
\State $\Delta E \gets |PosEyeFlow_{last} - PosEyeFlow|$
\If {$E > \epsilon$ OR $(\Delta E < \epsilon*0.1$ AND $!isPrecise$)} 
\State recalculateFacePose
\State $isPrecise \gets TRUE$
\Else
\State $isPrecise \gets FALSE$
\EndIf
\end{algorithmic}
%\end{lstlisting}
\end{algorithm}
%\vspace{-0.5cm}

$\epsilon$ refers to the spatial threshold and $E$ to the error in pixels between the current estimated eye positions and the position calculated in the last precise detection step. $\Delta E$ is the difference of $E$ between the current and the last head tracking frame. The threshold $\epsilon$ determines the trade-off between coarse but fast head pose estimation and precise but expensive tracking during user motion. Our system uses an empirically estimated $\epsilon$ of 3\% of the diagonal of the input image in pixels.

%\noindent
%\todoLipsum{1}
%%%%%%%%%%%%%%%%%%%%%%%%
%%% Performance Evaluation
%%%%%%%%%%%%%%%%%%%%%%%%
%\newpage
\section{Performance Analysis}
We compared the rendering performance of our system (AAUPR) to full featured UPR and FUPR (which does not require any head tracking). The performance measures of have been recorded on an HTC-M8 Android smartphone. The numbers in Table~\ref{timings} indicate mean values over 1000 frames for all conditions. \textit{Resolution} refers to the image resolution of the back camera, \textit{Frame Time} refers to the average time spend to render a single frame, and \textit{Tracking Time} refers to the time spent for head tracking in total over 1000 frames. The resolution of the back camera was set to 640x480 pixel in all conditions, and visual tracking was performed in the same environment for during all system measurements to provide the same number of visual features in all conditions. The front camera delivered new video frames at a maximum of 15 frames per second (which is the hardware limit of the camera). Please note that the head tracker runs in a separate thread.

\begin{table}[t!]
\vspace{-0.45cm}
		\centering
		\caption{Performance measured in milliseconds.}%\vspace{0.5cm}}
		\begin{tabular}{cccc}
%\hline
System & Resolution & Frame Time & Tracking Time \\
UPR   & 320x240  & 29.379 & 14.080 \\
AAUPR & 320x240  & 23.890 &  4.602   \\
FUPR  & 320x240  & 20.733 & 0.0  \\
\hline
UPR   & 640x480  & 42.860 & 30.094  \\
AAUPR & 640x480  & 28.706 & 13.363  \\
FUPR  & 640x480  & 20.733 & 0.0  \\
		\end{tabular}
		\label{timings}
\end{table}

%\todo{discuss performance measures}
%\todoLipsum{1}

%%%%%%%%%%%%%%%%%%%%%%%%
%%% Experiment
%%%%%%%%%%%%%%%%%%%%%%%%
\section{User Experiment}
\noindent
\textbf{Design.} 
We designed a repeated measures within-subjects study to compare the performance and user experience of different implementations of user-perspective rendering. Therefore, we introduced an independent variable rendering with four conditions: device-perspective rendering (DPR), user-perspective rendering (UPR), approximated user-perspective rendering (FUPR), adative user-perspective rendering (AAUPR). 

The task was a pointing task in which participants had to align the mobile device with a circular target area and touch the target area while looking through the view of the mobile device. The target area was only visible in the device view so that participants were forced to interact with the target area by using the different rendering conditions. Like Pucihar et al.~\cite{vcopivc2013evaluating} we are interested in the effect of the spatial distortion when looking through the device. Therefore, we do not show the live video during our experiment so that participants do not see their hands during interaction. The target area had a radius of 20mm based on the recommended size of ISO-9241~\cite{ISO9241} for button input. The viewpoints of the target areas were placed randomly. For each rendering condition participants performed 40 repetitions. Rendering was counterbalanced using a balanced Latin square. 

As dependent variables we measured task completion time (TCT) and error of each task, subjective workload was measured by the raw NASA TLX~\cite{hart1988development}, usability using the Single Ease Question (SEQ) \cite{Sauro:2009} and overall preference.

Sixteen participants (3 female, $\overline{X}=$30.7 (sd$=$3.5) years old) volunteered for the study. On a scale from one to five, five meaning best, the mean of self-rated AR experience was 3.3 (sd$=$1.4).

\textbf{Apparatus.} 
Initially, we planned to perform the experiment using our mobile implementation on an Android based smartphone (HTC-M8). However, due to the computational demands of the head tracking, the phone overheated and throttled the CPU during the pilot test after approximately 5 minutes in full UPR mode. All other conditions did not show this behaviour. However, since the phone didn't cool down fast enough, CPU throttling had an impact on all subsequent measurements. Therefore, we did not use a mobile phone in the study setup, but settled with a PC setup and a wired connection to a mobile display (Figure~\ref{fig:experiment}).

The apparatus consists of an installed touch screen and a handheld screen. The installed screen was a Dell S2340T multi-touch monitor of size 23" (506 x 287 mm) and was used for recording touch input of the user. The handheld touchscreen was a HTC M8 phone (5" screen, 109 x 61 mm) attached to a PC via USB and was used to achieve an AR view implementing the different rendering conditions. The circular target areas of the task were shown as augmentation registered on the installed screen. The augmentation was only visible when viewed through handheld AR device. The touch screen could be rotated to allow participants to comfortably work with the screen while standing. The head tracking and the tracking of the handheld screen was performed on the PC.

\textbf{Procedure.}
After an introduction and filling out a demographic questionnaire, measurements were taken to calibrate the systems of the rendering condition. We measured inter-pupillary distance and the distance between the handheld device and the participant's head to set up the fixed viewpoint for FUPR. To determine the distance participants were asked to hold the handheld device centered onto the touch screen at a distance of 15 cm. The distance was only calibrated once. Afterwards, participants familiarized themselves with the first rendering condition by performing the task until they felt comfortable using the system. Then, the measured tasks started. Participants were instructed to quickly point to the target area by performing one fluid natural hand motion to the area, where the target was expected. Participants were instructed to not move out of their initial position, but to turn their body to reach the target areas farther away from the center. This should simulate the movement of narrow work spaces, where the head position is not always ideal for FUPR.

For each task, participants first had to locate the target area by searching with the AR view. After locating the target area, participants touched the handheld device screen and then, with the same hand, the target area on the screen. The TCT was measured between the touch of the handheld device and the touch screen. Hence, TCT does not include search time for the target, but focuses only on the coordination of the hand as expected to be seen through the AR device. Error was recorded as the Euclidean distance between the detected touch point and the center of the target area. Participants received a distinct visual and audio confirmation, for either hitting or missing the target area. 

Participants performed 40 repetitions per rendering condition. After a rendering condition, participants filled out the NASA TLX and the SEQ. The next rendering condition started thereafter, following the same procedure. After finishing the last rendering condition and filling out NASA TLX and SEQ, participants filled out a preference questionnaire and a semi-structured interview was conducted. A session took approximately 30 minutes.

With 16 participants, four rendering conditions and 40 repetitions per rendering condition, there were a total of $16 \times 4 \times 40 = 2560$ trials.

\textbf{Hypotheses.}
Due to the nature of the pointing task we did not expect significant differences in task completion time. However, due to the spatial distortion of the view of the DPR condition, we expected DPR to have a significantly higher error rate than any other condition (\textbf{H1}). Due to the optimal compensation of the spatial distortion, UPR will have less errors than FUPR (\textbf{H2}). We hypothesize that our novel approach taken with AAUPR produces significantly less errors than FUPR (\textbf{H3}). %We also expected that our AAUPR implementation will be non-inferior to UPR (\textbf{H4}).  \todo{if testing for non-inferiority, we also should test AUPR for non-inferiority. actually aupr performed better than upr in all measurements! upr and aupr lie in the same range

\begin{table}[t!]
	\centering
	\resizebox{\columnwidth}{!}{
	\begin{tabular}{|c|c|c|c|c|}
		\hline
		       & \textbf{DPR} & \textbf{UPR} & \textbf{FUPR} & \textbf{AAUPR} \\\hline
		Time (s)     & 1.5 (0.8)   & 2 (1.4)     & 1.7 (1.1)   & 1.8 (1.1)   \\
		Error (pxl)  & 26.8 (14.4) & 17.3 (11.6) & 20.9 (12.5) & 15.9 (10.4) \\		
		TLX          & 52 (16.3) & 38.6 (14.6) & 44.6 (14) & 30.9 (14.4) \\
		SEQ          & 3.1 (1.6) & 4.75 (1.3) & 3.6 (1.5)  & 5.3 (1.7) \\
		Preference   & 0 & 2 & 3 & 8 \\
		\hline
	\end{tabular}
	}
	\caption{Study results. Mean and standard deviation of time and error, and SEQ and TLX results. Last row indicates the number of participants preferring the interface. Three participants did not state a clear preference, except not choosing DPR.}
	\label{Table:study_results}
\end{table}

\textbf{Results}
  The data was evaluated using a level of significance of 0.05. The data fulfilled sphericity and normality requirements and, therefore, was analyzed using ANOVA and post-hoc pairwise t-tests with Bonferroni correction. Questionnaire data was analyzed using a non-parametric Friedman tests followed by pairwise Wilcoxon signed rank tests with Bonferroni correction. The reported p-values have been Bonferroni corrected to reflect a significance level of 0.05. The statistical analysis was performed using the software R.

The ANOVA revealed a significant difference in error for the rendering condition (F(3,45)=12.26, p$<$0.001). Post-hoc tests revealed significant differences between DPR and UPR (p$<$0.001), DPR and AAUPR (p$<$0.001) and a weak significant difference between FUPR and AAUPR (p$=$0.06).

Friedman tests revealed significant differences in TLX (${\chi}^2(3)=20.01,p<0.001$) and SEQ (${\chi}^2(3)=20.59,p<0.001$). Post-hoc tests revealed significant differences for TLX between DPR and AAUPR (Z=3.11, p$<$0.005), FUPR and AAUPR (Z=2.64,p$<$0.05) and a near significant difference between DPR and UPR (Z=-2.53, p=0.053). Post-hoc tests revealed significant differences for SEQ between DPR and UPR (Z=2.63, p$<$0.05), DPR and AAUPR (Z=3.09, p$<$0.005) and FUPR and AAUPR (Z=3.04, p$<$0.01).

\ifdefined\includeFigures
\begin{figure}[t!]
\centering
	%%%% 
%	\subfigure[Accuracy]
	{\includegraphics[width=0.99\columnwidth]{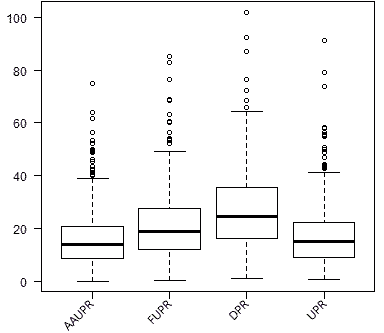}}	
	\caption{Study results. Boxplot of the error in pixels.}
	\vspace{-0.35cm}
\label{fig:results}
\end{figure}
\fi

\textbf{Discussion}
As hypothesized, the results of DPR were worst in all tested measurements. DPR had a significantly higher error than UPR and AAUPR due to the optimal spatial distortion of the rendered user-perspective view in these conditions. This is also reflected in the SEQ, which was rated significantly lower compared to UPR and AAUPR. In terms of TLX, DPR also had a significantly higher workload than AAUPR and a weak significant difference to UPR. However, we could not find a significant difference between DPR and FUPR and, therefore, could not replicate the findings of Pucihar et al.~\cite{vcopivc2013evaluating}. One possible explanation could be the nature of the task, that required interactions over a large distance which eventually requires updating the user's head pose relative to the device. Overall, we partially accept H1.
%One possible explanation could be the nature of the task of Pucihar, that encouraged participants to spend more time on pointing the target area, while our task required participants to spontaneously point to the target area. Therefore, we partially accept H1.

Interestingly, UPR did not perform significantly better than FUPR in any measurement. Therefore, we reject H2. We believe that the lack of performance comes from the implementation of the head tracking. During the experiments we noticed jitter in the head tracking, that might influenced the pointing accuracy. This also could explain the better performance of AAUPR, which did not suffer from the problem of continuous jitter, because the tracking rate was lower than the one of UPR. Hence, AAUPR performed significantly better than FUPR in terms of error, TLX and SEQ measurement. Thereby, we accept H3. \review{Note that the head tracking could be implemented more stable using head mounted fiducials. However, we are aiming at a mobile and self contained system, why we have implemented head tracking based on visual face tracking.}

In terms of user preference shown in the last row of Table~\ref{Table:study_results} three participants did not clearly prefer any user perspective rendering. However, they were clear in indicating that they did not prefer DPR. For the preference results of the other 13 participants we performed an exact binomial test comparing to chance (0.25) and found a significant difference in preference for the AAUPR interface (p$<$0.05). This is in line with the results of the study, indicating an advantage of AAUPR over FUPR. The significant preference over UPR also underlines the advantage of AAUPR in terms of more stable, discrete tracking updates. 

\review{To be able to record user interactions we performed the experiment on a large touch screen. However, the interaction space is often much larger which requires the user to move the AR display much more around (see Figure\ref{fig:overview} for a real life example). Since our system is designed to  compensate user motion, we believe that interactions in spaces larger than the one used in our experiment will lead to similar results or to a favorable bias towards AAUPR.}

%\ifdefined\includeFigures
%\begin{figure}[t!]
%\centering
%	%%%% 
%	{\includegraphics[width=0.99\columnwidth]{img/tmp_column.jpg}}	
%	\caption{TLX}
%\label{fig:tlx}
%\end{figure}
%\fi
%\newpage
\section{Conclusion and Future Work}
We have presented a system for user perspective rendering on handheld devices which supports large interaction spaces. Our system does not perform 3D head tracking in each frame, instead we measure motion over time to trigger updates. This reduces the overall computational demands of the system. During our experiment we furthermore noticed that less updates of the user's head pose provides more stable renderings on mobile phones. We believe this is the reason why users prefer our approach over continuous user perspective rendering. %This being said, one could also potentially improve the stability of monocular head tracking algorithms by temporal or spatial filtering.

While our system reduces the impact of tracking failure, erroneous head tracking still impacts the performance of our system to some degree. Therefore, future work needs to further investigate 3D head tracking on mobile devices. In addition, we will further optimize the number of necessary updates of user's head pose. In this regard, we will incorporate further information available to the system at run-time, such as the state of the application or the current task to perform or the estimated distance between the current pose and an anticipated future pose of the device.  

%Our system is designed for the current generation of mobile devices. However, faster and more reliable head tracking approaches, in combination with more powerful devices will make our adaptive solution for user perspective rendering less important. Therefore, future work needs to further investigate head tracking on mobile devices. Still, 

%\todoLipsum{1}

%\todoLipsum{1}

%\ifdefined\includeFigures
%\begin{figure*}[t!]
%\centering
%	%%%% 
%	\subfigure[]
%	{\includegraphics[width=0.65\textwidth]{img/figure_overview.png}}	
%	%\subfigure[]
%	%{\includegraphics[width=0.65\textwidth]{img/figure_overview.png}}	
%	\caption{Applications}
%\label{fig:applications}
%\end{figure*}
%\fi

%f specified like this the section will be ommitted in review mode
\acknowledgements{
This work was funded by a grant from the Competence Centers for Excellent Technologies (COMET) 843272 and the EU FP7 project MAGELLAN (ICT-FP7-611526).
}

%\newpage
\balance
\bibliographystyle{abbrv}
%%use following if all content of bibtex file should be shown
%\nocite{*}
\bibliography{bib}
\end{document}